\documentstyle[preprint,aps,psfig]{revtex}
 \tightenlines
 \begin{document}
 \title{ From Storage and Retrieval of  Pulses to Adiabatons}
 \author{Tarak Nath Dey and G. S. Agarwal}
 \address{Physical Research Laboratory,
 Navrangpura, Ahmedabad-380 009, India}
 \date{\today}
 \maketitle
 \begin{abstract}
 We investigate whether it is possible to store and retrieve the intense probe
 pulse from a $\Lambda$-type homogeneous medium of cold atoms. Through numerical
 simulations we show that it is possible to store and retrieve the probe pulse
 which are not necessarily weak. As the intensity of the probe pulse
 increases, the retrieved pulse remains a replica of the original pulse, however
 there is overall broadening and loss of the intensity. These effects can be
 understood in terms of the dependence of absorption on the intensity of the
 probe. We include the dynamics of the control field, which becomes especially
 important as the intensity of the probe pulse increases. We use the theory of
 adiabatons [Grobe {\it et al.} Phys. Rev. Lett. {\bf 73}, 3183 (1994)] to
 understand the storage and retrieval of light pulses at moderate powers.
 \end{abstract}

 \newpage
 \section{Introduction}
 Photons are ideal carriers of information. However it is extremely
 difficult to store photons for a long time and to retrieve
 them from a medium. The basic principle of photon storage is based on ultraslow group
 velocity of the light\cite{Hau,Kash,Bud,Sch,Hemmer}. The later is made possible by
 electromagnetically induced transparency(EIT) \cite{Harris} where an external control field
 is used to make an atomic medium transparent even for a frequency  near  atomic
 resonance. Under such conditions, a probe pulse at a particular frequency and polarization can
 propagate with a substantially reduced group velocity. Experimental and theoretical
 studies \cite{Liu,Cerboneschi,Phillips}show that the ultraslow group velocity of the pulse
 causes spatial
 compression of a probe pulse that has a spatial length of several kilometers in
 free space.
 This spatial compression leads to localization of the pulses within the atomic medium.
 Within the spatially localized pulse region, atoms are in a  state of superposition
 characterized
 by the amplitudes and phases of the control and probe fields. Therefore, the induced
 transparency is associated with a substantial reduction of
 the group velocity of the probe pulse which is formed due to  a coupled field -spin
 excitation, called dark state polaritons \cite{Phillips}.
 Switching off of the external control field converts
 the dark state polaritons into purely atomic coherence which is confined within the medium.
 By switching on the external control field at a later time, the atomic coherence can be
 transferred back into the radiation field. The regenerated radiation field can
 be an exact replica of the original one
 \cite{Juzeliunas,Zibrov} . Therefore, the signal pulse information
 can be stored and retrieved by switching off and on  the external optical field
 \cite{Liu,Phillips,Juzeliunas,Zibrov,Matsko}.
 The mode of switching can be adiabatic\cite{Phillips} as well as nonadiabatic \cite{Matsko}.
 The past studies\cite{Liu,Phillips,Juzeliunas,Zibrov,Matsko} on this problem have been carried out in the
 linear regime, where the probe pulse is much weaker than control field.
 Now the question is, whether it is possible to store and retrieve an intense
 probe pulse, in which case, the non-linearity of the medium with respect to the probe pulse
 amplitude becomes important? In this paper we address this question. We  characterize the probe pulse
 propagation for different types of pulses in a homogeneously broadened
 $\Lambda$- type medium. The paper is organized as follows: In Sec. II, we obtain the
 Maxwell-Bloch equations that governs the propagation dynamics of the optical pulse at
 moderate power inside the $\Lambda$-type homogeneous medium.
 We include the radiative decay in simulations.
 The results of our numerical simulations which also includes the dynamical evolution of the
 control field are presented in Sec. III.
 In Sec. IV, we show how the results of numerical simulations can be understood in terms of
 the adiabaton theory of Grobe {\it et al.}\cite{Grobe}.

 \section{DYNAMICAL EQUATIONS for PULSES PROPAGATION AT MODERATE POWERS}
 We consider an atomic system where relevant atomic transitions are taken in the
 configuration as shown in the Fig. 1. wherein, one of the
 transition is coupled to a external control field $\rm{E}_c$ and the other transition is
 coupled to a probe field $\rm{E}_p$. We define all fields as
 \begin{equation}
 \rm{\vec{E}(z,t)}=\vec{\cal {E}}(z,t)e^{-i(\omega t-k z)}~+~c.c.~,
 \end{equation}
 where $\vec{\cal {E}}$~is a slowly varying envelop, $\omega$~is the carrier frequency
 and $k$ is the wave number of the field.  Here we assume the frequencies of
 the carrier waves coincide with the frequencies of the corresponding atomic
 transitions $\omega_{_{1}}=\omega_{_{{13}}}$ and $\omega_{_{2}}=\omega_{_{12}}$.

 We are using the Maxwell-Bloch Equation to describe the propagation dynamics of
 light pulses through the atomic vapour. In the slowly varying envelope
 approximation the temporal and spatial evolution of the field envelops is
 governed by
 \begin{eqnarray}
 \frac{\partial g}{\partial z}+ \frac{\partial g}{\partial
 ct}&=&i\eta \rho_{_{13}}\nonumber\\
 \frac{\partial G}{\partial z}+ \frac{\partial G}{\partial
 ct}&=&i\eta \rho_{_{12}}~,
 \end{eqnarray}
 where $2g=2\vec{d}_{13}\cdot\vec{\mathcal{E}}_p/\hbar
 ~{\textrm {and}}~ 2G=2\vec{d}_{12}\cdot\vec{\mathcal{E}}_c/\hbar$ are the Rabi
 frequencies of the probe and drive fields, $\vec{d}_{13}$ and $\vec{d}_{12}$ are
 the dipole moments, $\rho_{_{13}}$ and $\rho_{_{12}}$ are the density matrix
 elements of the corresponding atomic transitions, and c is the velocity of light
 in vacuum.
 The propagation constant is taken to be equal for all transitions and
 is given by $\eta=3\lambda^2N\gamma/8\pi$. It depends on the density
 of the atomic vapor N and the atomic transition wavelength $\lambda$.

 Within the rotating wave approximation the interaction between the atoms and the
 fields can be describe by the Bloch equations in a rotating frame
 \begin{eqnarray}
 \dot{\rho}_{_{11}}&=&-2(\gamma_1+\gamma_2)\rho_{_{11}}+i G \rho_{_{21}}+i g
 \rho_{_{31}}-i G^{*} \rho_{_{12}}-i g^{*}
 \rho_{_{13}}~,\nonumber\\
 \dot{\rho}_{_{22}}&=&  2\gamma_2\rho_{_{11}}+i G^* \rho_{_{12}}  - i G
 \rho_{_{21}}
 ~,\nonumber\\
 \dot{\rho}_{_{12}}&=&-[\gamma_1+\gamma_2]\rho_{_{12}}+i G
 \rho_{_{22}}+ig \rho_{_{32}}-iG\rho_{_{11}},\\
 \dot{\rho}_{_{13}}&=&-[\gamma_1+\gamma_2]\rho_{_{13}}
 +iG\rho_{_{23}}+ig\rho_{_{33}}-ig\rho_{_{11}},\nonumber\\
 \dot{\rho}_{_{23}}&=& iG^*\rho_{_{13}}-ig\rho_{_{21}}~.\nonumber
 \end{eqnarray}
 Here $\gamma$'s govern the radiative decay of the state $|1\rangle$; we further assume
 the equality $\gamma_1=\gamma_2=\gamma/2$.
 These Bloch equations are to be supplemented by the population conservation law
 \begin{equation}
 \rho_{_{11}}+\rho_{_{22}}+\rho_{_{33}}=1
 \end{equation}
 The solution of Eqs.(2)-(3) gives the complete
 evolution of the atom-field system at any instant of time.
 The analytical solution of the Maxwell-Bloch equations not known though under
 special conditions some solutions are known \cite{Eberly}.
 Therefore we study the pulse-propagation problem  only numerically.

 \section{Numerical Simulations}
 \subsection{Pulse Propagation: Effect of nonlinearities}
 We solve the propagation problem numerically for a homogeneously broadened gas
 of cold $\Lambda$-atoms in the travelling window frame of reference:
 $\tau=t-z/c$,$~\zeta=z$.
 We consider initial probe pulses of two different shapes  given by

 \[ g(0,\tau) = \left\{ \begin{array}{ll} g^0e^{-\left(\frac{\tau-\tau_0}{\sigma}\right)^2} &
 \mbox{~Gaussian pulse}\\
 g^0[{\rm sech}(\frac{\tau-\tau_0}{\sigma})+ f\times{\rm
 sech}(\frac{\tau-\tau_1}{\sigma})] & \mbox
 {~Sech pulse.}
 \end{array}
 \right. \]
 Here, $g^0$ is the real constant characterizing the peak amplitude of the Rabi
 frequency before the pulse enter the homogeneous medium, $\sigma$ is the band
 width of the input pulse and $\tau_i$ determines the number of peaks as well as
 their position.
 The initial condition for the atomic system is to be taken as
 $\rho_{33}(\eta\zeta/\gamma,0)=1$,
 with all other density matrix elements equal to zero. In order to appreciate
 the effect of nonlinearities,
 we first consider the control field as a continuous wave $G(0, \gamma \tau)\equiv
 {\rm constant}$ $[(G/\gamma)^2=10]$. We work under the condition of
 electromagnetically induced transparency, Figures 2(a), 2(b), 2(c) and 2(d) display
 the propagation of two types of probe
 pulses inside the medium. From  Figs. 2(a) and 2(c), we see
 that the weaker probe pulse would  propagate without any significant absorption and broadening
 inside the medium. The Figs. 2(b) and 2(d)  show that the intense
 probe pulse suffers absorption and broadening. The shape of the pulse remains almost identical
 to the input pulse. This behavior of the intense probe
 pulse can be explained with help of steady state probe absorption spectra.
 In Fig. 3, we show the behavior of the probe absorption  as a function of the
 probe detuning when the control field is on resonance. It is
 clear from the Fig. 3 that, increase of the probe field intensity results in
 the increased absorption of the probe for a given frequency in the
 neighborhood of the frequency satisfying two photon resonance condition.
 Note that the transparency window that appears in the
 absorption spectrum has a finite bandwidth which depends on the intensities of
 the  control and probe fields. The width of the transparency window becomes
 narrow when the probe field intensity is increased while keeping the control
 field intensity constant.
 Therefore, the condition for distortionless pulse propagation is that the
 bandwidth of the probe pulse should be contained within the transparency window of
 the EIT medium. If the pulse becomes too short, or its spectrum too broad relative
 to the transparency window of the EIT medium, absorption and higher order
 dispersion need to be taken into account. The group velocity dispersion of the
 medium causes the broadening of the probe pulse.

 \subsection{Storage and Retrieval at Moderate powers}
 The light pulse propagating in a homogeneously broadened $\Lambda$-type medium,
 suitably driven by another control field, can be stooped and later released in
 a controlled way \cite{Liu,Phillips,Kochar,Agarwal}.
 The ultra slow group velocity of the light is the main key issue of the
 ``light-storage" technique. In particular, changing the control field intensity
 results in changing the group velocity of the light pulse.
 The smooth switch off and on of the control field is made possible by gradually
 varying the intensity of the control field with respect to time. Therefore,
 the switching off and on of the control field can be modelled by
 a super-Gaussian shape given by
 \[ G(0,\tau)~=~G^0[1~-~e^{-\left(\frac{\tau-\tau_2}{\sigma'}\right)^{\alpha}}]~~
 \begin{array}{ll} \alpha=4 & \mbox{adiabatic switching}\\
                  \alpha=100 & \mbox{nonadiabatic switching.}
 \end{array}
 \]
 In Fig. 4(a), we consider the adiabatic switching of the control field .
 Switching off of the control field  give rise to the absorption of the probe pulse
 when the entire probe pulse is inside the medium. The group velocity
 of the probe pulse is reduced to zero  and its
 propagation is stopped by switching off the control field.
 The stored probe pulse can be retrieved by switching on the
 control field. The time difference between switching off and on is
 dependent on the life time of the atomic coherence between the state $|2\rangle$ and
 $|3\rangle$. As seen from  Figs. 4(b) and
 4(d), for weaker probe pulse,
 the shape of the retrieved pulse is same as the original one
 because the width of the probe pulse spectrum is very much less than width of
 the EIT window. Therefore, almost perfect storage and retrieval of light is
 possible by adiabatic switching of the control field as pointed out by
 Fleischhauer {\it et al.}\cite{Phillips}.
 When the
 probe field intensity is large, we observe from Figs. 4(c) and 4(e),that the retrieved
 probe pulse suffers absorption as well as broadening because of narrowing of
 the width of EIT window. Remarkably, the probe pulse can be
 retrieved even for probe that is not necessarily weak.
 We also present Figure 5 to show the behavior of the atomic
 coherence $\rho_{_{32}}$ as a function of retarded time, which depends on the behavior
 of the input probe pulse. In presence of control field,
 the temporal shape of the atomic coherence $\rho_{_{32}}$ is same as the shape
 of the input probe pulse. Remarkably enough the $\Lambda$ system leads to storage and
 retrieval even at moderate powers of the probe field.
 \subsection{Nonadiabatic Results}
 Scully et al.\cite{Matsko} have shown that for any switching time of the control
 field, an almost perfect storage and retrieval of weak probe pulse is possible.
 We extend storage and retrieval of the probe pulse in the nonlinear regime.
 The results have been shown in Fig. 6 for an intense probe pulse and
 nonadiabatic switching of the control field. For both adiabatic and nonadiabatic
 switching, the retrieved intense probe pulse is same as original
 one. However, there is overall broadening and loss in intensity of the retrieved
 probe pulse. The Fig. 6(b) shows that the drop in intensity ratio of the output
 to input probe pulse as function of the input pulse intensity.

 \subsection{Dynamical Evolution of the Control field}
 The dynamical evolution of the control field  becomes important when the
 intensities of the control and probe fields are of comparable strength.
 Figure 7 depicts time evolutions of the control field at different distances.
 It is evident from this figure that a dip and a bump develops in the amplitude
 of the control
 field as its propagates through the medium. The shape of the bump and dip in
 the control field  depends on the initial shape of the input probe pulse at the
 entry surface of the medium. The propagation dynamics of the dip of the control
 field and broadened probe field together can be understood in terms of adiabaton
 pair\cite{Grobe}.
 \section {Adiabaton Theory and Its Relation to Light Storage}
 In a remarkable paper Grobe {\it et al.} \cite{Grobe} discovered what they
 called as adiabatons. These are the pulse pairs which are generated in a
 $\Lambda$-system under conditions of adiabaticity. We show the deep connection of
 the problem of storage and retrieval of pulses to the theory of adiabatons.
 The control field is switched on before the probe field, to keep the system
 in the dark state and which is an essential condition for the adiabaton formation.
 Under conditions of negligible damping, the
 response of the medium then can be very well approximated by the solutions
 \begin{eqnarray}
 \rho_{13} &\approx& \frac{i}{V}\frac{\partial}{\partial
 \tau}\left(\frac{g}{V}\right)\nonumber\\
 \rho_{12} &\approx& \frac{i}{V}\frac{\partial}{\partial
 \tau}\left(\frac{G}{V}\right)\\
 \rho_{32} &\approx&- \frac{g G}{V^2}~\nonumber,
 \end{eqnarray}
 provided the following adiabacity relation is satisfied by the two field:
 \begin{equation}
 G\frac{\partial g}{\partial \tau} - g\frac{\partial G}{\partial \tau} \ll V^3,
 \end{equation}
 where $V^2=(G^2 + g^2)$. By inserting solution(5) into the Maxwell equation(2),
 we obtain a pair of coupled nonlinear wave equations
 \begin{eqnarray}
 \frac{\partial g}{\partial \zeta} &=& -\frac{\eta}{V}\frac{\partial}{\partial
 \tau}\left(\frac{g}{V}\right)\nonumber\\
 \frac{\partial G}{\partial \zeta} &=& -\frac{\eta}{V}\frac{\partial}{\partial
 \tau}\left(\frac{G}{V}\right).
 \end{eqnarray}
 Note that the radiative decay constant does not play any role in the above
 two equations. These two, one dimensional PDE are nonlinearly coupled
 through the variable V. With the help of equation (7), one can easily show that V does
 not depend on $\zeta$, implying that V is remains constant during the
 propagation.
 \begin{equation}
 V\left(\frac{\eta \zeta}{\gamma}, \gamma\tau\right)=V\left(0, \gamma\tau\right)
 \end{equation}
 Thus the conservation law would imply that V
 any change in probe field is compensated by a corresponding in
 change the control field. Analytical solutions of equation (7) can also
 be obtained by changing the variable $\tau$ to $z(\gamma\tau) \equiv
 \frac{1}{\gamma^2}\int_{-\infty}^{\gamma\tau}V^2(0, \gamma\tau)d(\gamma\tau)$:
 \begin{eqnarray}
 g\left(\frac{\eta \zeta}{\gamma}, \gamma\tau\right) &=& V(0, \gamma\tau) F_g
 \left[z(\gamma\tau)-
 \frac{\eta \zeta}{\gamma}\right]\nonumber\\
 G\left(\frac{\eta \zeta}{\gamma}, \gamma\tau\right) &=& V(0, \gamma\tau) F_G
 \left[z(\gamma\tau)-
 \frac{\eta \zeta}{\gamma}\right].
 \end{eqnarray}
 $F_g[x]=g[0, z^{-1}(x)]/V[0, z^{-1}(x)]$ and $z^{-1}(x)$ denotes the
 inverse function of z. We have chosen the initial fields strong enough to ensure
 the formation of an adiabaton pair. The input fields g and G are chosen such that
 V is constant after a certain time T; hence, for $\tau\geq  T$ the
 integral $z(\gamma\tau)$ can be analytically performed. First we take the control field
 of a constant amplitude $(G/\gamma=3.16)$ and the probe field as Gaussian pulse,
 we then obtain (for $\tau\geq T$)
 \begin{eqnarray}
 g\left(\frac{\eta\zeta}{\gamma},\gamma\tau\right)&=&\frac{\surd[{g^0}^2 e^{-
 \frac{2(\gamma\tau-\gamma\tau_0)^2}{\sigma^2}}+{G^0}^2]}{\surd[{g^0}^2
 e^{-\frac{2(\gamma\tau-\gamma\tau_0-\frac{\eta\zeta}{\gamma {G^0}^2})^2}{\sigma^2}}
 +{G^0}^2]}
 g^0 e^{-\frac{(\gamma\tau-\theta_0)^2)}{\sigma^2}}\nonumber\\
 G\left(\frac{\eta\zeta}{\gamma},\gamma\tau\right)&=&\frac{\surd[{g^0}^2 e^{-
 \frac{2(\gamma\tau-\gamma\tau_0)^2}{\sigma^2}}+{G^0}^2]}{\surd[{g^0}^2
 e^{-\frac{2(\gamma\tau-\gamma\tau_0-\frac{\eta\zeta}{\gamma {G^0}^2})^2}{\sigma^2}}
 +{G^0}^2]}
 G^0~~.
 \end{eqnarray}
 Next the control filed is taken as a super Gaussian shaped pulse and the probe field as
 Gaussian pulse as before. The solutions of equation (7) for both these cases, are superimposed on the
 numerical results obtain from density matrix formalism in Figure 7.
 It is remarkable that, the solutions of equations (7) obtain under adiabatic approximation,
 matches extremely well with the numerically solutions of the density matrix formalism containing
 seven nonlinear equations.
 The adiabatic calculation of atomic coherence $\rho_{32}$ of the equation(5) is indistinguishable from
 the density matrix formalism as shown in Fig. 5 \cite{tarak}.
 It is evident from the temporal profiles of the control and probe at different propagation distances
 that a dip and a bump develops in the control
 field intensity as it propagates through the medium. Figure 7 unambiguously
 confirms that the adiabaton pair ( consisting of the dip in the pump and
 broadened probe) travels loss-free distances which exceed the weak probe
 absorption length(here typical value of $\eta\zeta/\gamma=2400$)by several orders
 of magnitude with an unaltered shape. A gradual decrease of the control field
 intensity $(G/\gamma)^2$ results in a corresponding decrease of the intensity of
 the probe pulse $(g/\gamma)^2$ maintaing the constancy of $V^2$, as shown in Fig.
 7(b). Therefore, it is clear that when the control field intensity becomes zero
 the probe pulse gets stored inside the medium. The reverse phenomena of the
 retrieval of the probe pulse is achieved by gradual increase of the control field
 intensity.  The fact that the numerical results from density matrix formalism
 matches extremely well with the adiabatic approximation clearly indicates that
 the storage and retrieval of light can be understood in terms of the adiabaton pair
 propagation.

 \section{Conclusions}
 We have investigated and answered in affirmative, the possibility of storage
 and retrieval of moderately intense probe pulses in a system with $\Lambda$ configuration.
 The propagation of two type probe pulses is first analyzed numerically, using
 Maxwell Bloch equation.
 It was found that the intense probe
 pulses are absorbed and broadened due to the non-linear dependence of
 susceptibility in the medium. In addition,  for the constant
 control field intensity, the width of EIT window  becomes narrower with a
 simultaneous increase of the probe field intensity. Therefore, when the
 intensity of the probe pulse
 is large, the non-linear effect of the medium becomes more important.
 For larger intensity probe pulses, storage and retrieval is  possible for
 both adiabatic as well as nonadiabatic switching of the control field. We
 studied the dynamics of the control field and found that the dip of the control
 field and broadened probe pulse  propagate together inside the medium as an
 adiabaton pair. We  show that retrieval probe pulse is the part of
 propagating adiabaton pair. Hence, the storage and retrieval of light can be
 clearly understood in terms of adiabaton pair propagation.
 The same was also studied by an adiabatic approximation. Remarkably,
 the result of the both cases are indistinguishable from each other in the wide
 parameters range studied here.

 \section*{Appendix: Numerical Integration Procedure}
 The integration of the equation is performed by the fourth order Runge-Kutta
 method\cite{press}.In our numerical calculation, sampling point along $\tau$
 and $\zeta$ direction are $4\times 10^6$ and $2\times 10^6$ respectively.
 We prefer the parallezation of the sequential code because of the
 large number of sampling points which requires large execution time. The
 sequential code can be easily parallelized if each iteration is
 independent of the other, that is no variables that are written in some
 iteration will be read and/or written in another iteration. But in our code
 each iteration is dependent on the previous iteration. Thus for our flow
 dependence case, the parallelization of the code is
 difficult and we do the parallelization manually by using OPEN MP
 directives in RS6000 in IBM machine.
 
 \newpage
 \begin{figure}
 \vspace*{5 cm}
 \centerline{\begin{tabular}{c}
 \psfig{figure=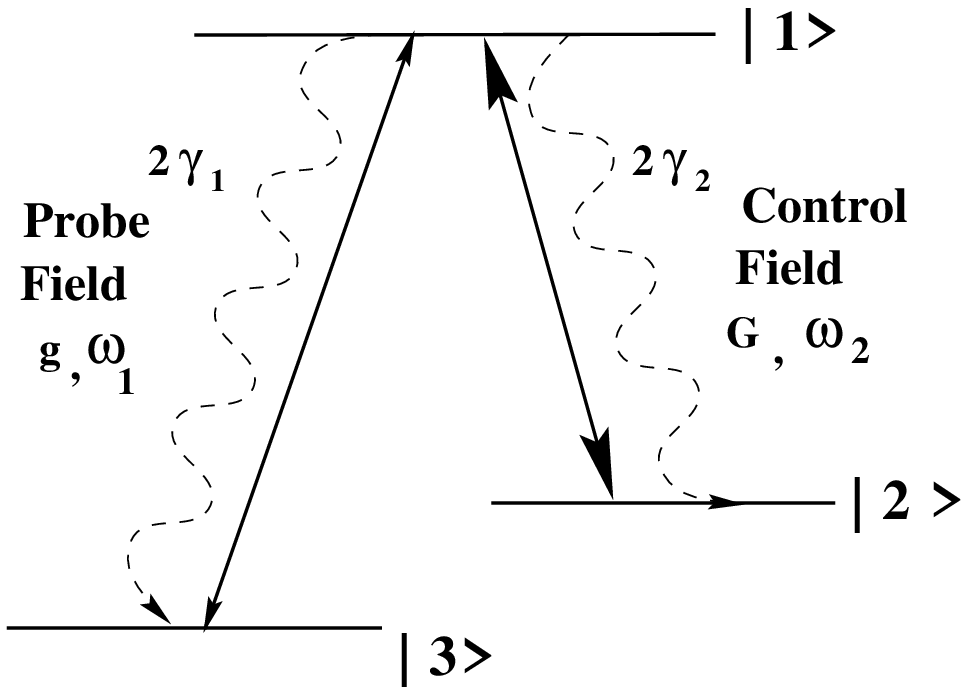,height=10.cm,width=12.0cm}
 \end{tabular}}
 \caption{ Three-level $\Lambda$-type medium resonantly coupled to a control
 field with Rabi frequency $G$ and probe field $g$.}
 \end{figure}
 \newpage
 \begin{figure}
 \centerline{\begin{tabular}{cc}
 \psfig{figure=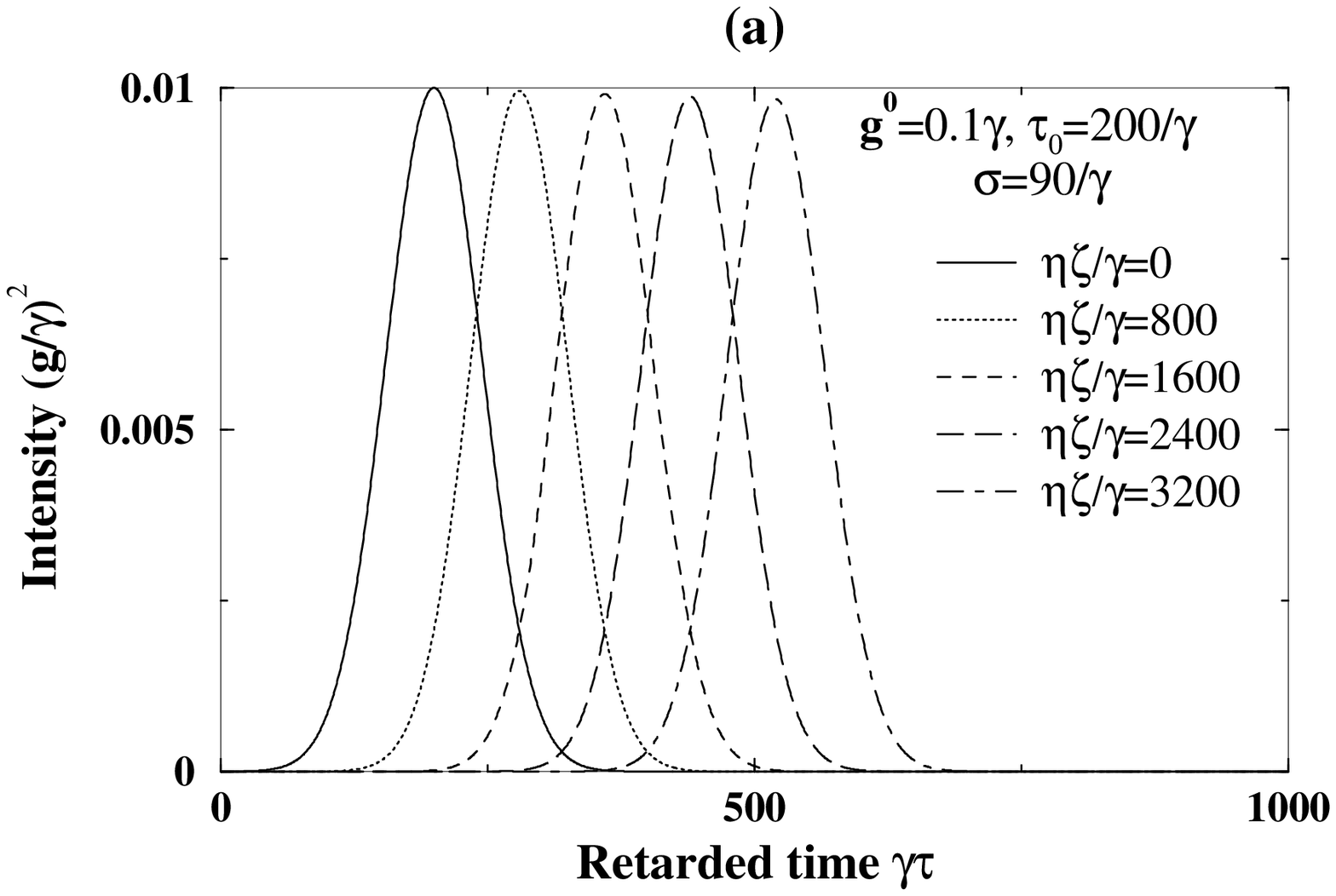,height=8 cm,width=9 cm}&
 \psfig{figure=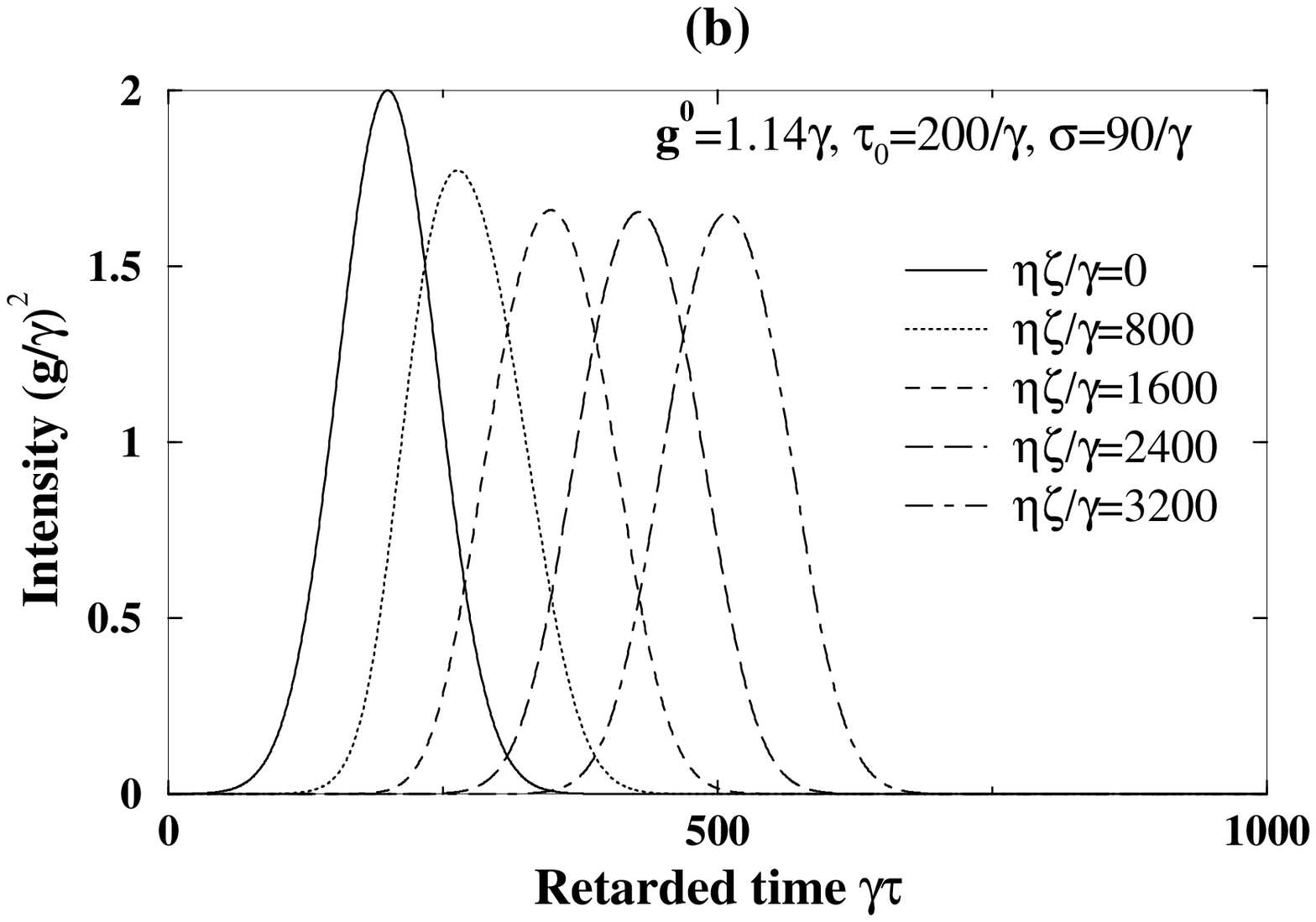,height=8 cm,width=9 cm}\\\\\\
 \psfig{figure=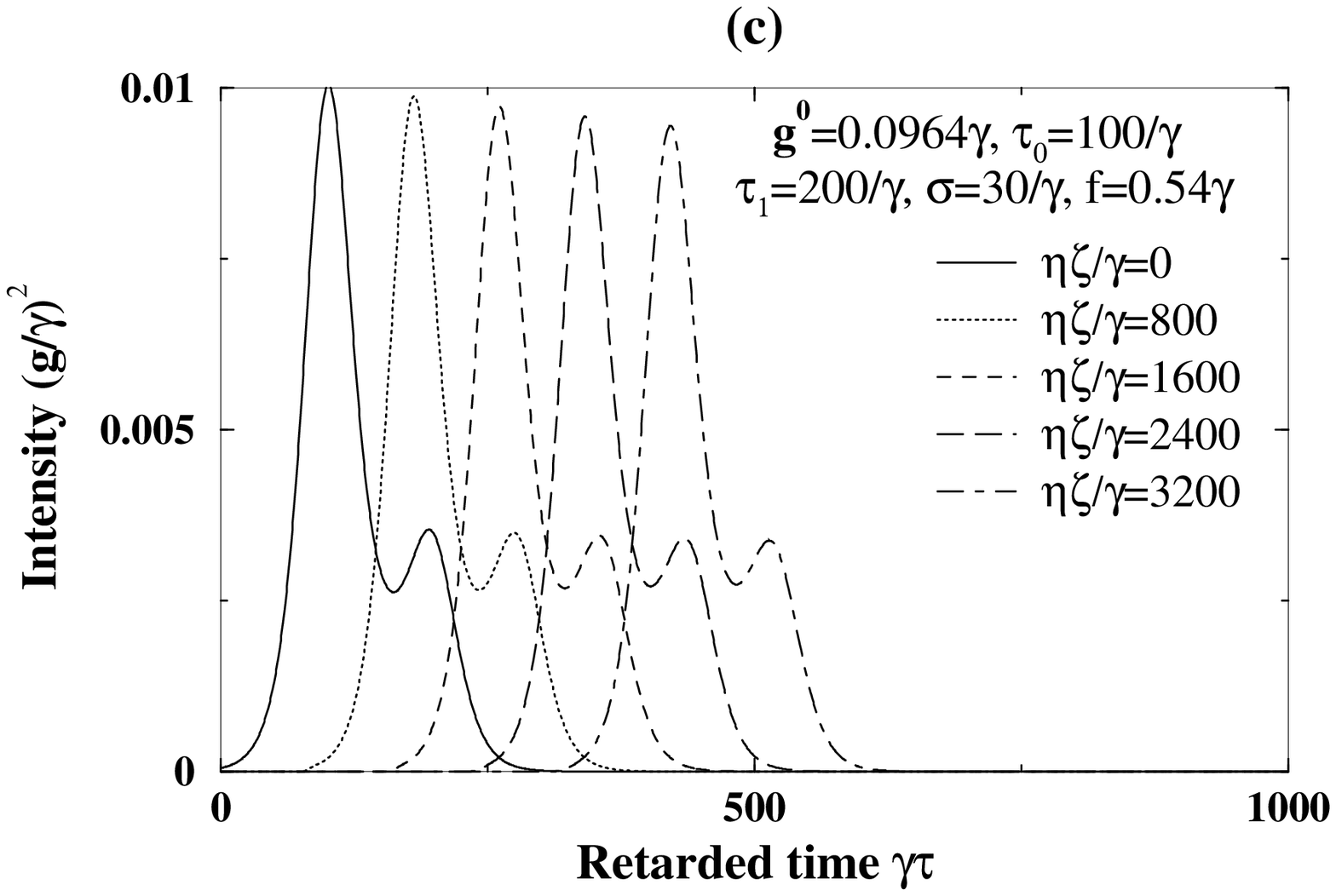,height=8 cm,width=9 cm}&
 \psfig{figure=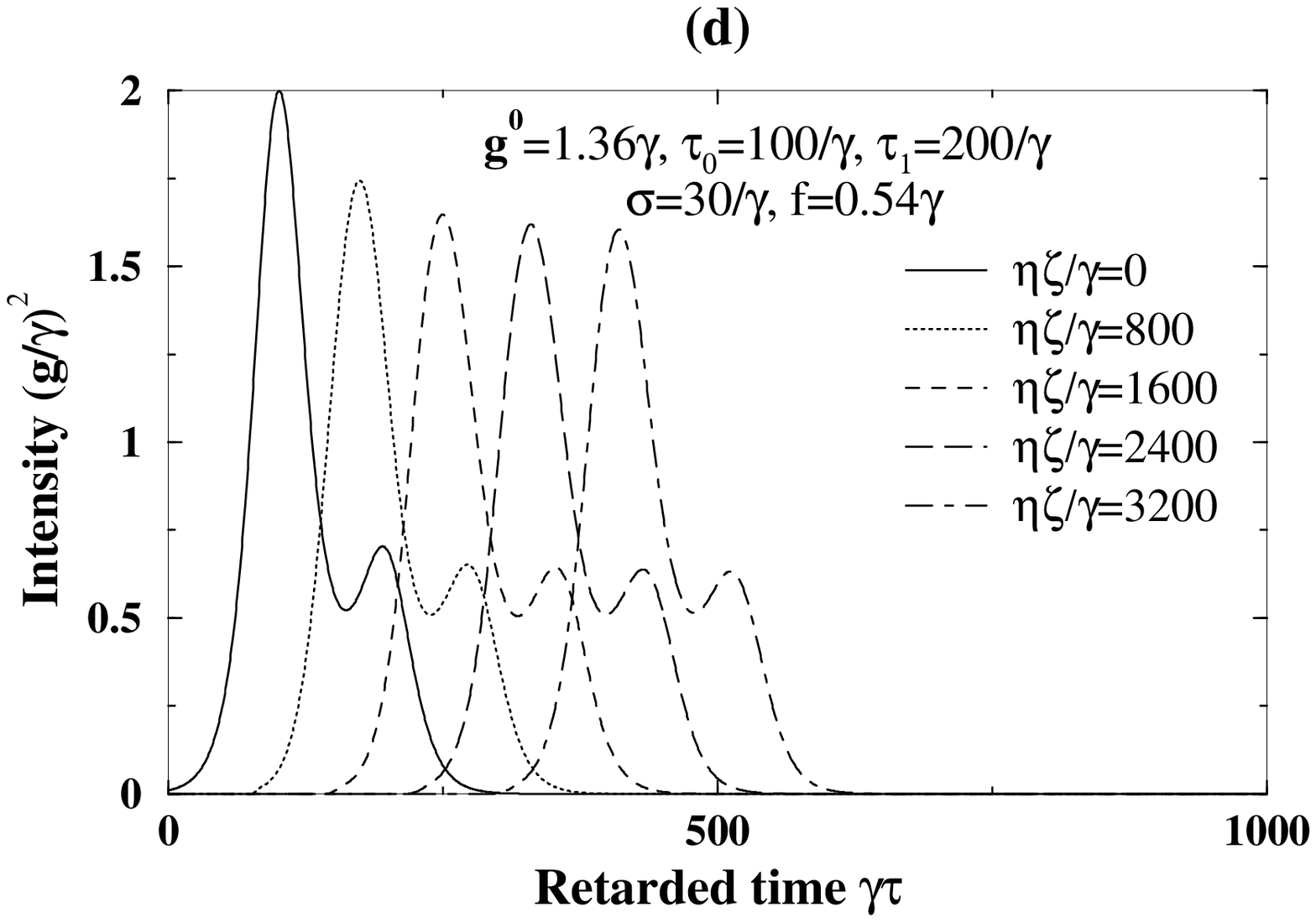,height=8 cm,width=9 cm}
 \end{tabular}}
 \caption{ The probe field intensity in the medium is plotted against retarded
 time at different propagation distances within the medium. Fig (a) and (c) show
 the probe pulse propagation with non-diminishing amplitude, for small intensities.
 Fig (b) and (d) depict the broadening and loss of intensity in case of an intense
 probe pulse case. In all the cases the control field is taken as CW$(G=3.16\gamma)$.}
 \end{figure}
 \newpage
 \begin{figure}
 \vspace*{5 cm}
 \centerline{\begin{tabular}{c}
 \psfig{figure=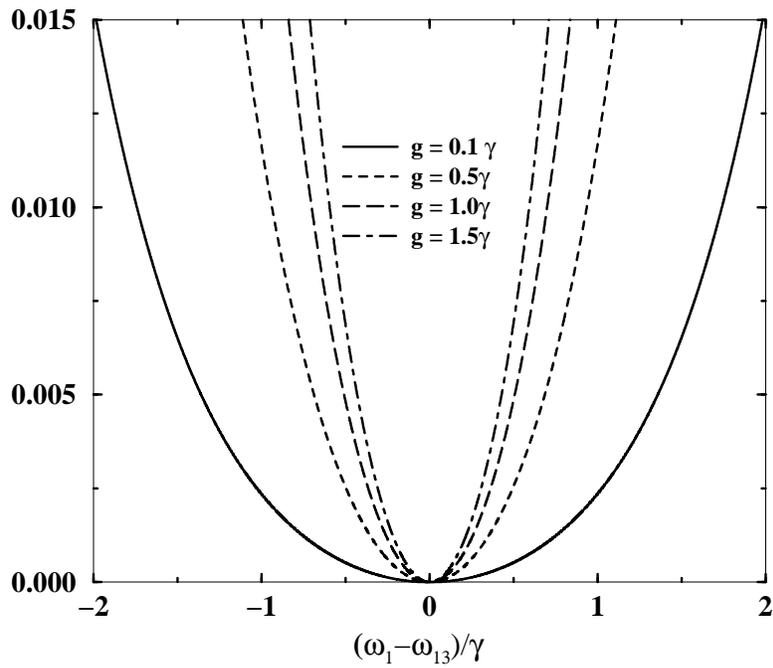,height=10.cm,width=12.0cm}
 \end{tabular}}
 \caption{Imaginary parts of susceptibility $[\chi_{_{13}}\hbar \nu / N|d_{_{13}}|^2]$ at a probe
 frequency $\omega_{_{1}}$ in the presence of control field as a CW ${\rm G}=3.16\gamma$. The width
 of the transparency window decreases with increase in the intensity of the probe field. }
 \end{figure}
 \newpage
 \begin{figure}
 \centerline{\begin{tabular}{c}
 \psfig{figure=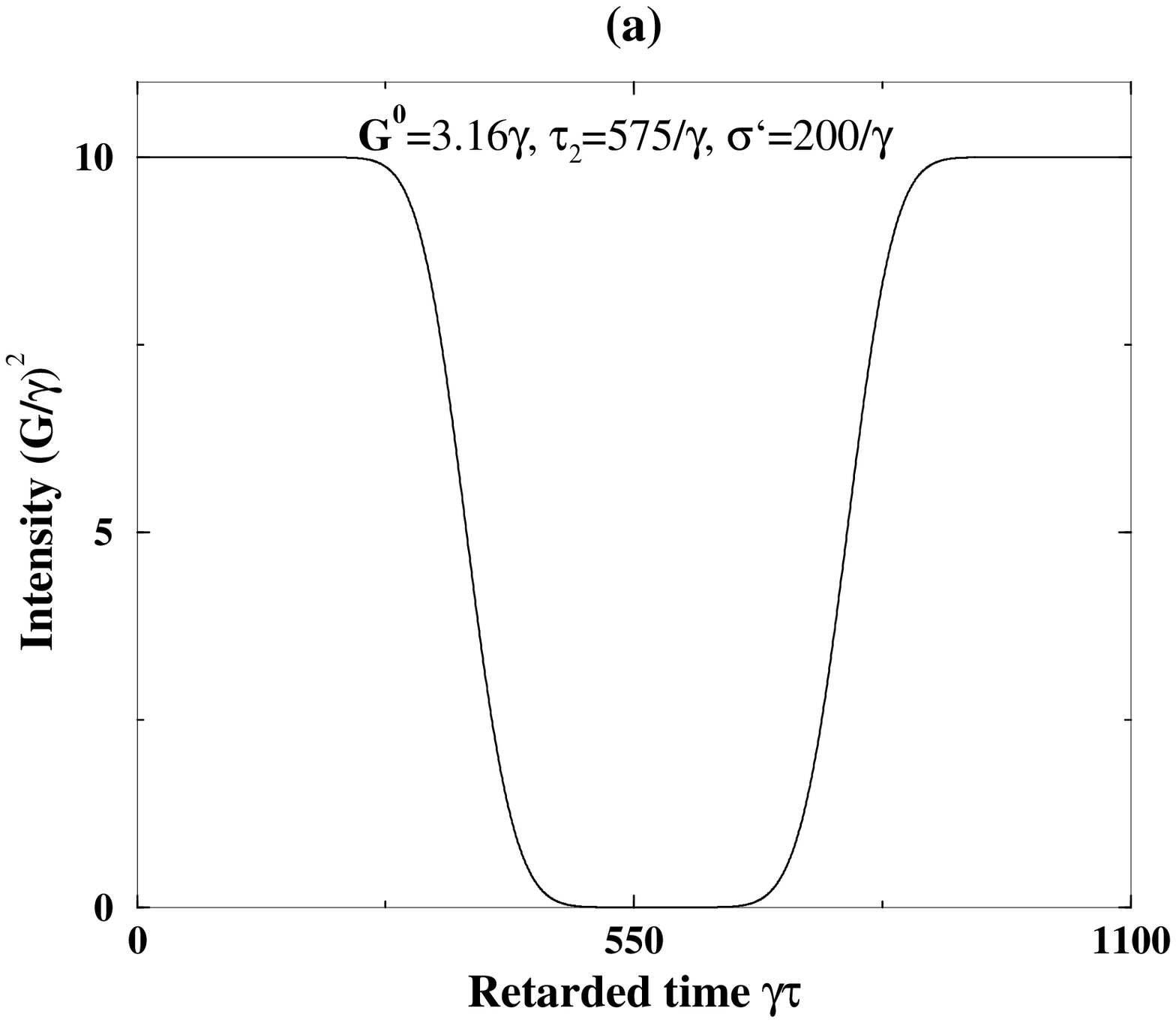,height=5.85cm,width=8cm}
 \end{tabular}}
 \end{figure}
 \begin{figure}
 \centerline{\begin{tabular}{cc}
 \psfig{figure=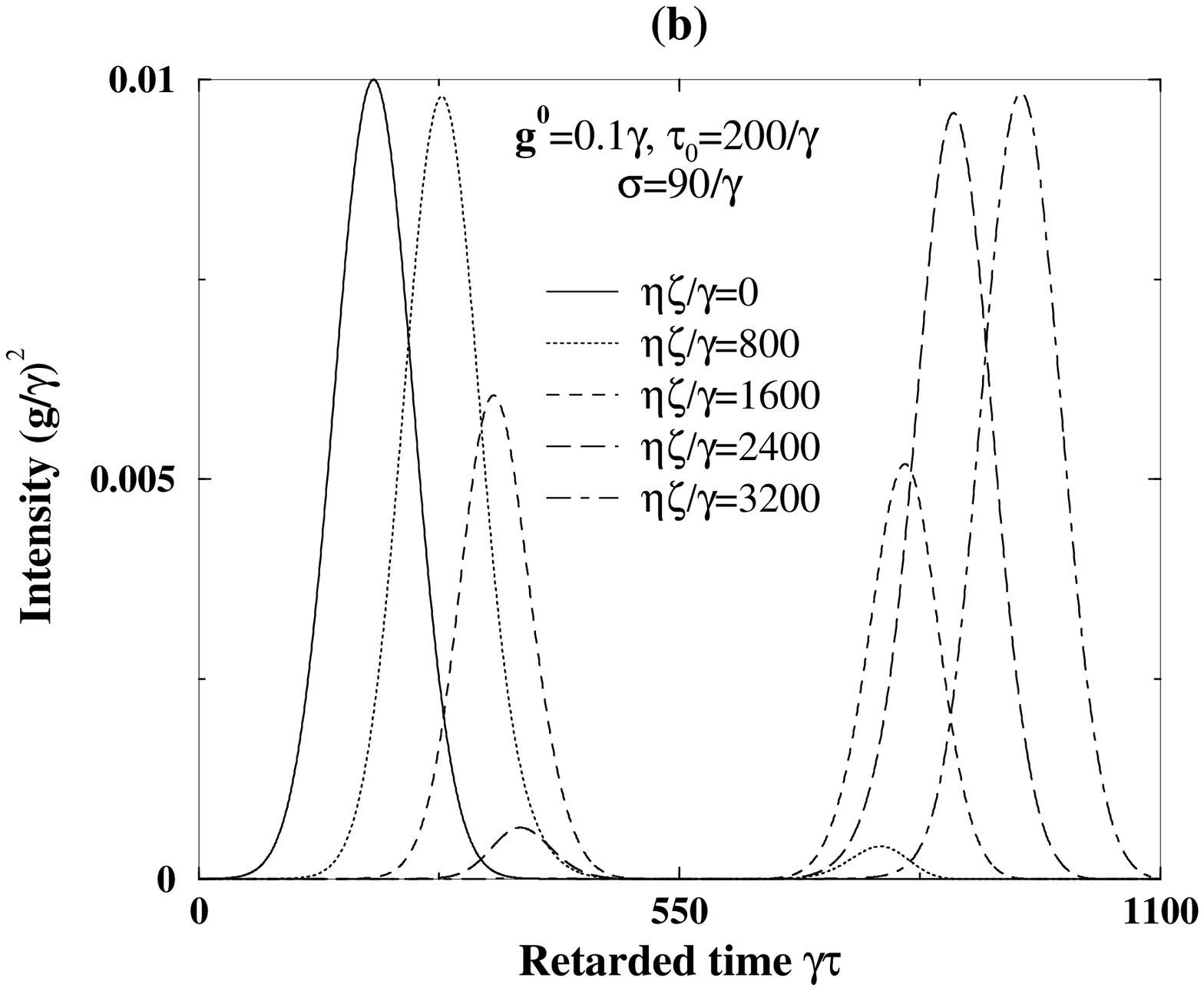,height=5.85 cm,width=8cm}&
 \psfig{figure=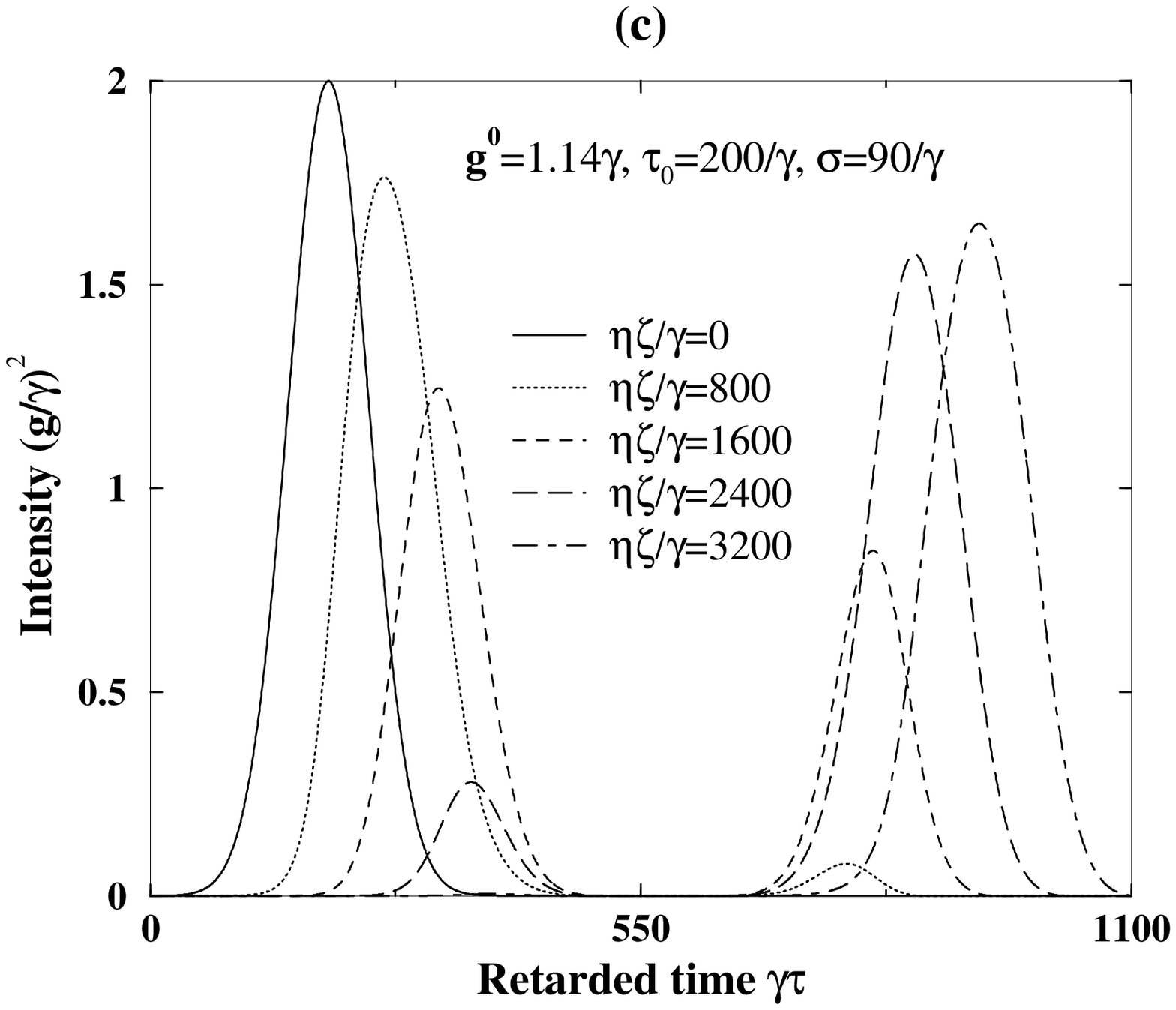,height=5.85 cm,width=8cm}\\
 \psfig{figure=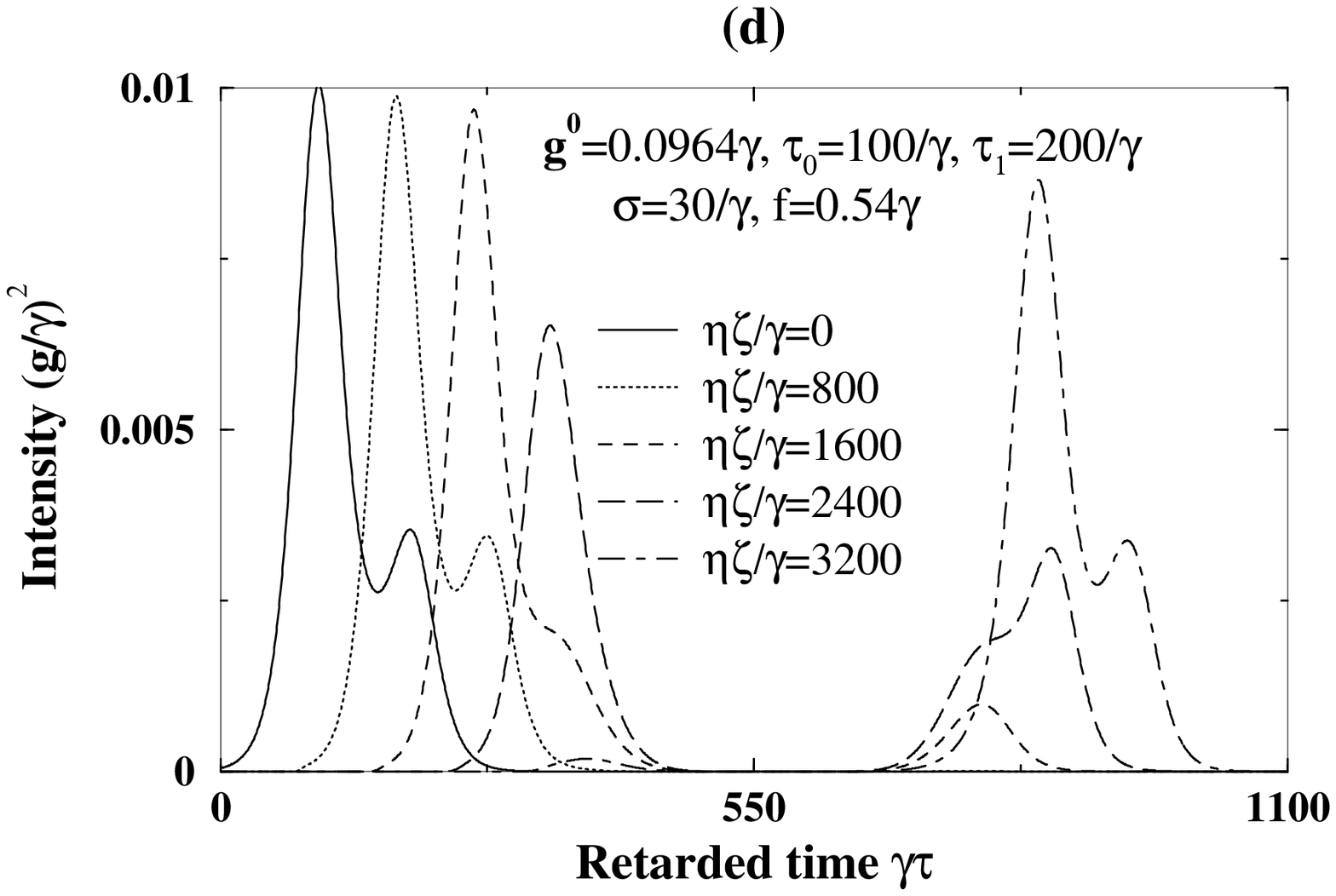,height=5.85 cm,width=9cm}&
 \psfig{figure=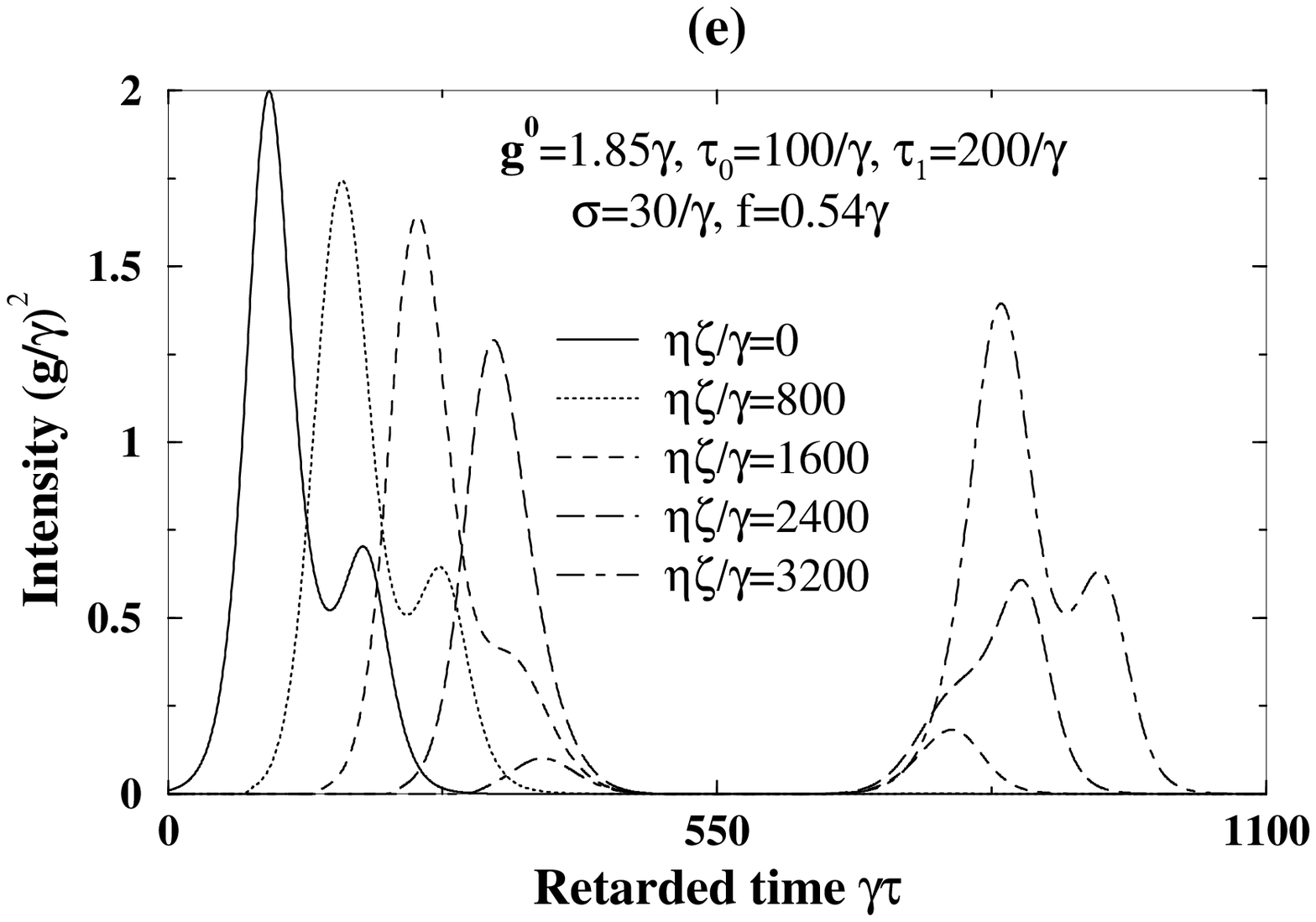,height=5.85 cm,width=9cm}
 \end{tabular}}
 \caption{{(a) shows the intensity of the control field as a function of
 retarded time at the entry surface of the medium at $\zeta=0$. Switching mode
 of the super Gaussian control field is adiabatic. The frames (b) and (d) show the time evolution
 of the weaker probe pulse at different propagation distances; and the frames (c) and
 (e) depict the time profile of the intense probe pulse at different propagation distances.}}
 \end{figure}
 \newpage
 \begin{figure}
 \centerline{\begin{tabular}{cc}
 \psfig{figure=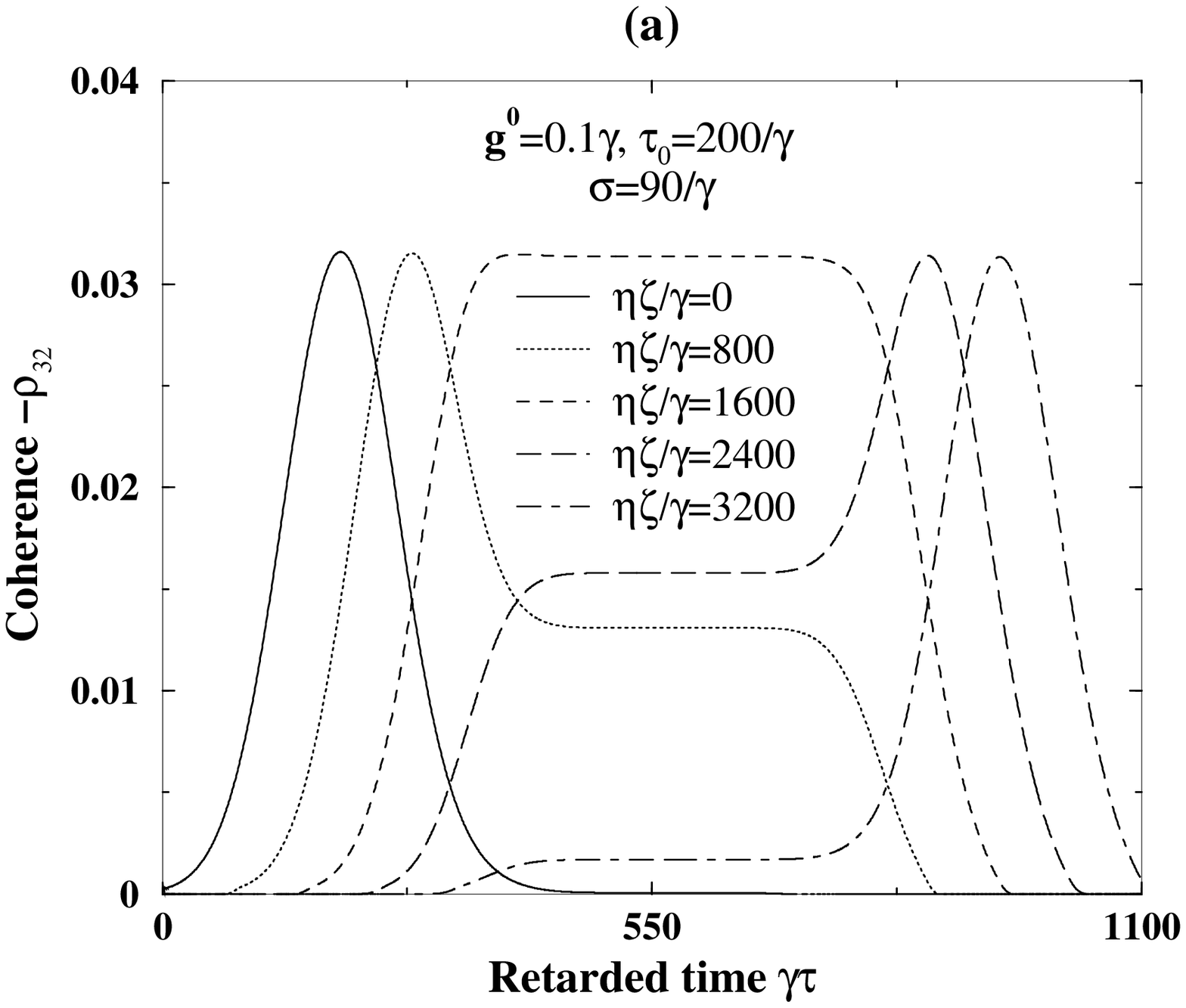,height=8cm,width=9cm}&
 \psfig{figure=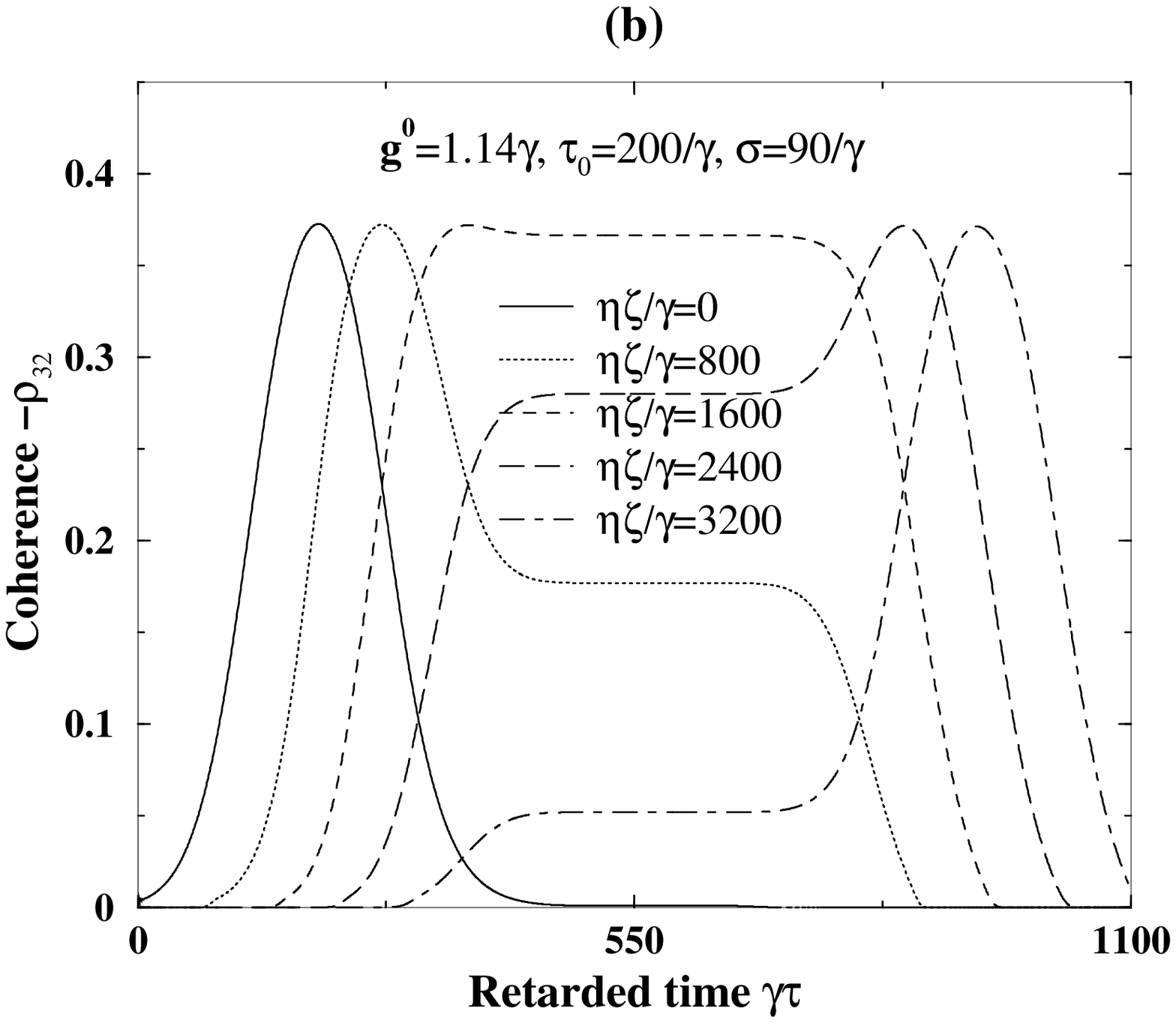,height=8cm,width=9cm}\\\\
 \psfig{figure=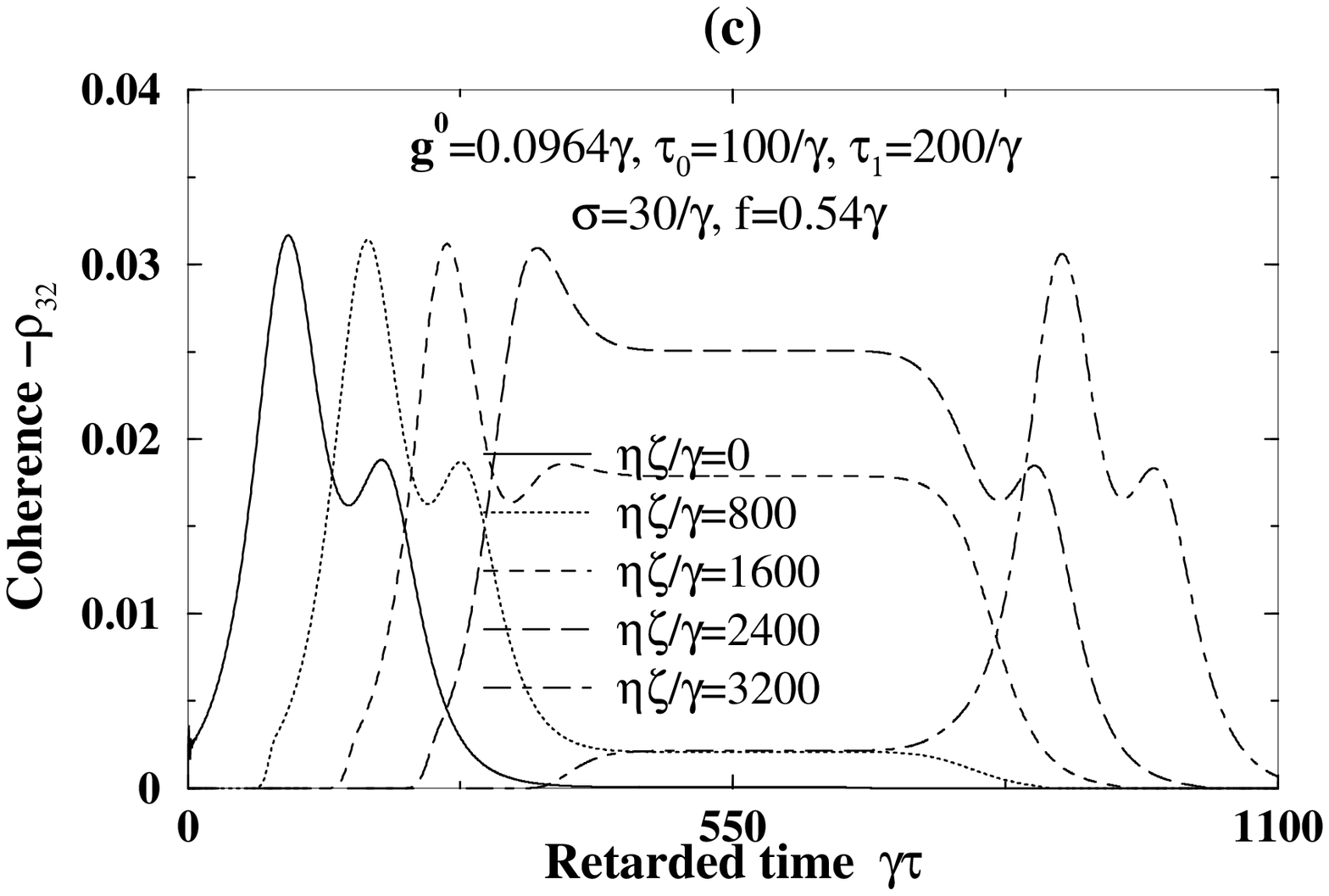,height=8cm,width=9cm}&
 \psfig{figure=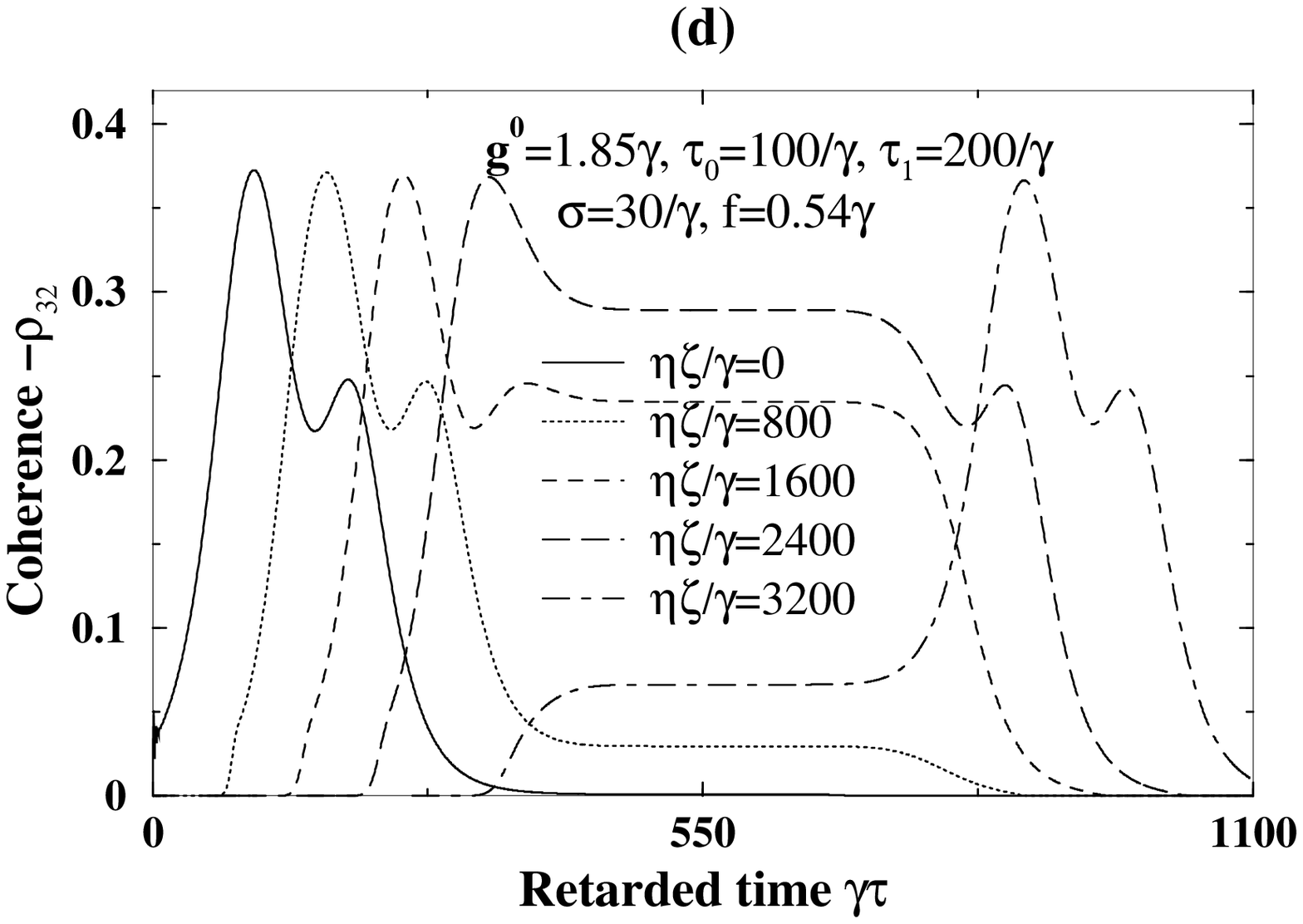,height=8cm,width=9cm}
 \end{tabular}}
 \caption{Fig. (a)-(d) shows the temporal profile of atomic coherence $-\rho_{32}$
 against retarded time at different propagation distances.}
 \end{figure}
 \begin{figure}
 \centerline{\begin{tabular}{cc}
 \psfig{figure=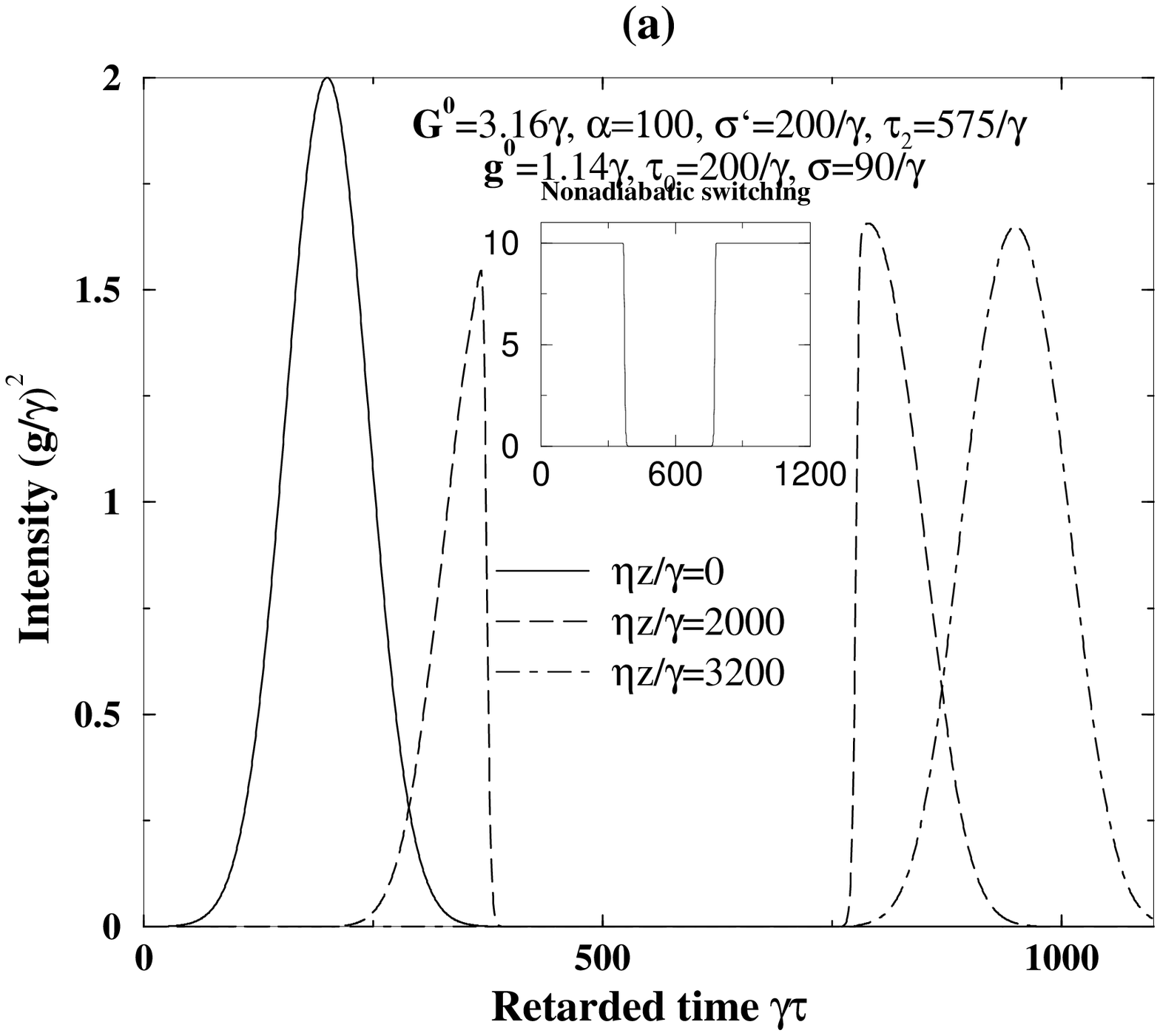,height=6.6cm,width=8cm}&
 \psfig{figure=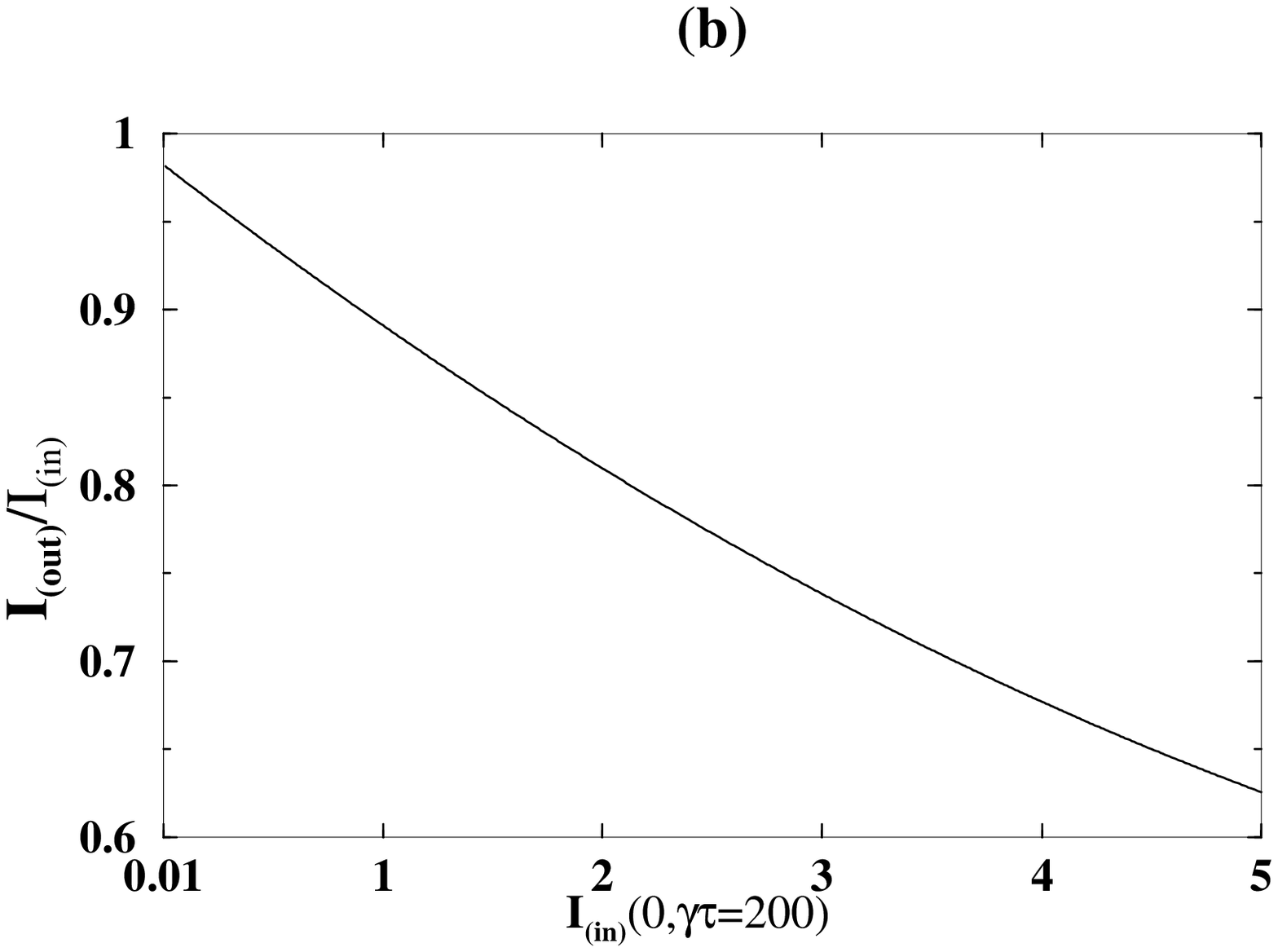,height=6.6cm,width=8cm}
 \end{tabular}}
 \caption{(a) shows storage and retrieval of intense probe pulse even for
 nonadiabatic switching of the control field. (b)drop in intensity ratio of the probe
 retrieved to the input
 pulse as a function of input probe intensity for the case of non adiabatic
 switching of the control field
 ; the ${\rm I}_{out}$ is measured at $\eta
 \zeta/\gamma=3200$ and $\gamma \tau =1000$. }
 \end{figure}
 \begin{figure}
 \centerline{\begin{tabular}{cc}
 \psfig{figure=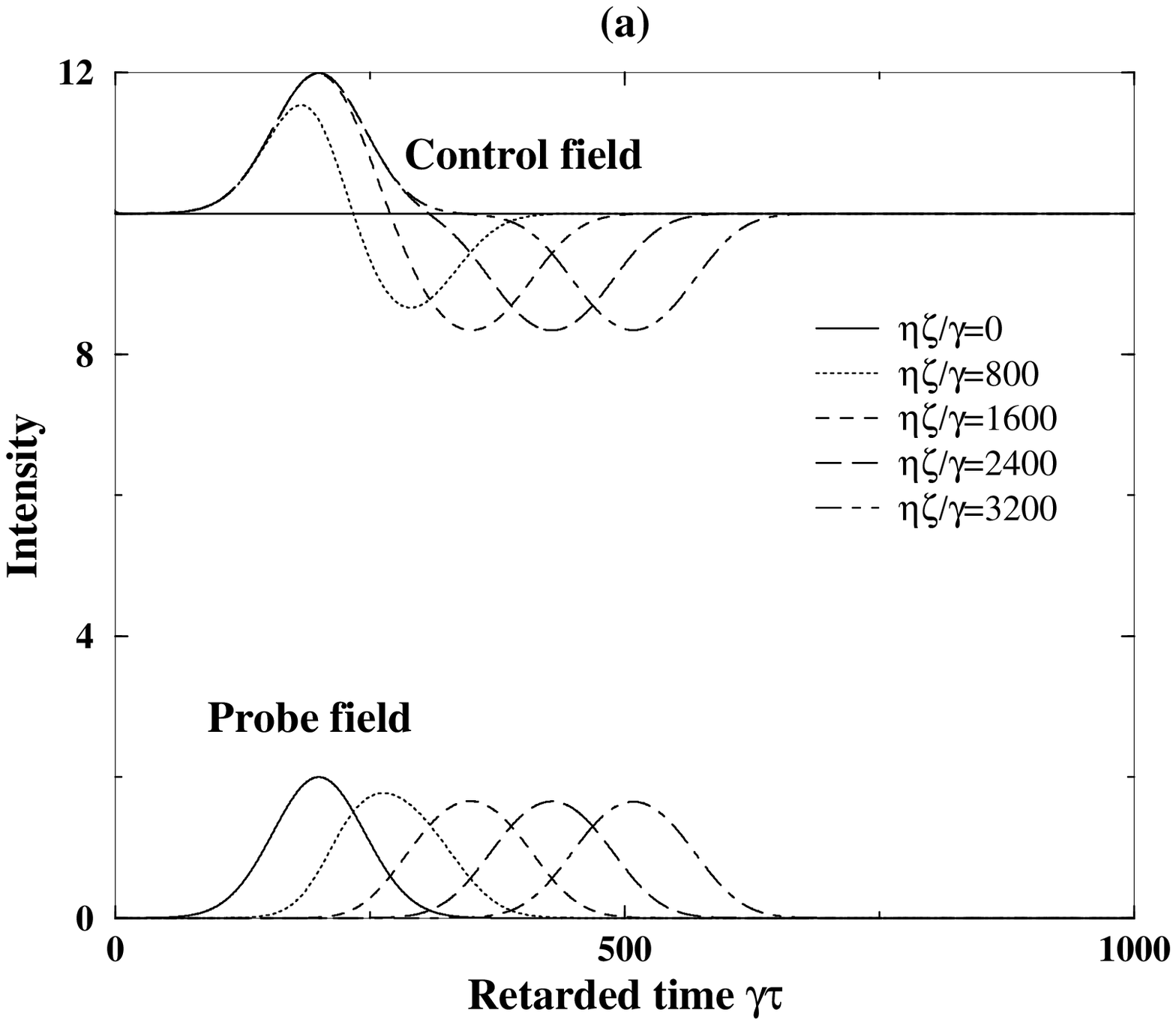,height=6.6 cm,width=8.0cm}&
 \psfig{figure=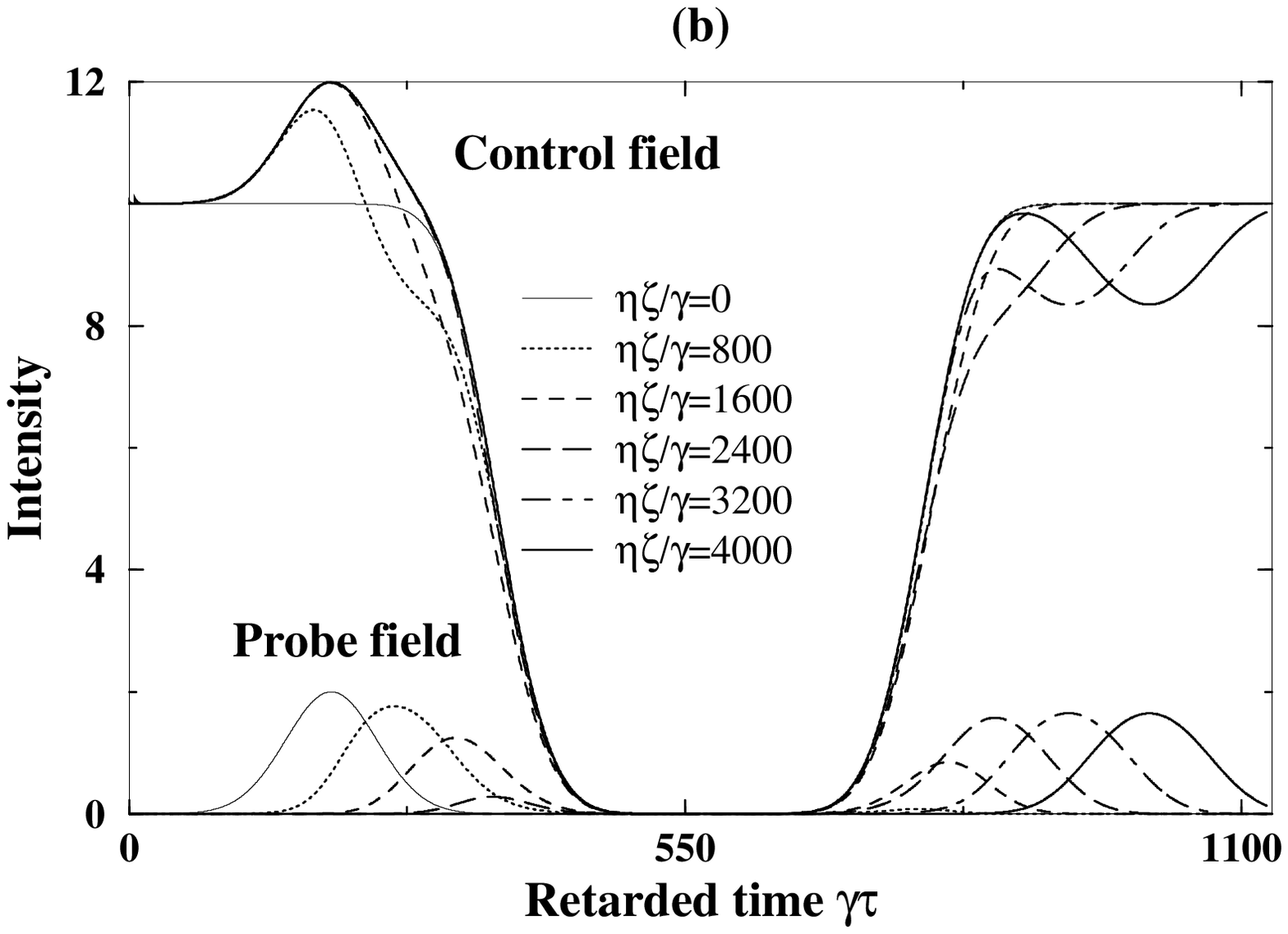,height=6.6 cm,width=8.0cm}
 \end{tabular}}
 \caption{(a) and (b) shows temporal profiles of the control $(G/\gamma)^2$ and
 probe field $(g/\gamma)^2$ at different
 propagation distances within the medium. In Fig. (a) the input control field is a
 CW. In Fig (b) the input control field is a Super-Gaussian shape with parameter
 $\tau_2=575/\gamma, \sigma\prime=200/\gamma$. The common parameter of the above two
 graphs  are chosen as: $G^0=3.16\gamma, g^0=1.14\gamma, \tau_0=200/\gamma, \sigma=90/\gamma$.
 The results of simulations using Maxwell-Bloch equations are indistinguishable from the
 results based on adiabaton theory. }
 \end{figure}

\begin{thebibliography}{99}

 \bibitem{Hau} L. V. Hau, S. E. Harris, Z. Dutton, and C. H. Behroozi, Nature
 (London) {\bf 397}, 594 (1999).
 \bibitem{Kash} M. M. Kash, V. A. Sautenkov, A. S. Zibrov, L. Hollberg, G. R.
 Welch, M. D .Lukin, Y. Rostovtsev, E. S. Fry, and M. O. Scully, Phys. Rev. Lett.
 {\bf 82}, 5229 (1999); A. B. Mastko, O. Kocharovskaya, Y. Rostovtsev, A. S.
 Zibrov, M. O. Scully, Adv. Mol. Phys. {\bf 46}, 191 (2001).
 \bibitem{Bud} D. Budker, D. F. Kimball, S. M. Rochester, and V. V. Yashchuk,
 Phys. Rev. Lett. {\bf 83}, 1767 (1999).
 \bibitem{Sch}O. Schmidt, R. Wynands, Z. Hussein, and D. Meschede, Phys. Rev.
 A {\bf 53}, R27 (1996).
 \bibitem{Hemmer} A. V. Turukhin, V. S. Sudarshanam, M. S. Shahriar, J. A.
 Musser, B. S. Ham and P. R. Hemmer, Phys. Rev. Lett. {\bf 88}, 023602 (2002).
 \bibitem{Harris}S. E. Harris, Physics Today, {\bf 50}, 36 (1997).
 \bibitem{Liu} C. Liu, Z. Dutton, C. H. Behroozi, and L. V. Hau,
 Nature (London) {\bf 409}, 490 (2001).
 \bibitem{Cerboneschi} E. Cerboneschi, F. Renzoni, and E. Arimondo, J. Opt. B
 {\bf 4}, s267, (2002).
 \bibitem{Phillips} D. E. Phillips, A. Fleischhauer, A. Mair, R. L. Walsworth, and
 M. D. Lukin, Phys. Rev. Lett. {\bf 86}, 783
 (2001); M. Fleischhauer and M. D. Lukin, Phys. Rev. Lett. {\bf
 84}, 5094 (2000); M. Fleischhauer and M. D. Lukin, Phys. Rev. A. {\bf 65}, 022314
 (2002).
 \bibitem{Juzeliunas} G. Juzeliunas, and H. J. Carmichael, Phys. Rev. A {\bf 65},
 021601(R) (2002).
 \bibitem{Zibrov} A. S. Zibrov, A. B. Matsko, O. Kocharovskyaya, Y. V.
 Rostovtsev, G. R. Welch, and M. O. Scully, Phys. Rev. Lett. {\bf 88}, 103601
 (2002).
 \bibitem{Matsko}A. B. Matsko, Y. V. Rostovsev, O. Kocharovskaya, A. S. Zibrov,
 and M. O. Scully, Phys. Rev. A {\bf 64}, 043809 (2001).
 \bibitem{Kochar} O. Kocharovskaya, Y. Rostovtsev, and M. O. Scully,
 Phys. Rev. Lett. {\bf 86}, 628 (2001).
 \bibitem{Agarwal} G. S. Agarwal and T. N. Dey,  accepted in JMO; G. S. Agarwal,
 T. N. Dey , and S. Menon, Phys. Rev. A, {\bf 64}, 053809 (2001).
 \bibitem{Eberly}J. H. Eberly and V. V. Kozlov, Phys. Rev. Lett. {\bf 88},
 243604 (2002); A. Rahman and J. H. Eberly, Phys. Rev. A {\bf 58}, R805 (1998).
 \bibitem{Grobe} R. Grobe, F. T. Hioe, and J. H. Eberly, Phys. Rev. Lett. {\bf
 73}, 3183 (1994).
 \bibitem{tarak} It should be noted that the adiabatic aproximation starts breaking down
 when  control field is switched off; however the created coherence servives.
 \bibitem{press}W. H. Press, S. A. Teukolsky, W. T. Vellerling and B. P.
 Flannery,
 {\it Numerical Recipes in Fortran},( Cambridge University Press, 1986).
 \end{thebibliography}
 \end{document}